\documentclass{ws-procs961x669}
\usepackage{float}
\usepackage[caption=false]{subfig}
\usepackage{url}

\begin{document}
\title{Advantages of Including\\
Globular Cluster Millisecond Pulsars\\
in Pulsar Timing Arrays}

\author{M. Maiorano$^*$, F. De Paolis, A. A. Nucita and A. Franco}
 
\address{Department of Mathematics and Physics ``Ennio De Giorgi'',\\
University of Salento, Via Arnesano, I-73100 Lecce, Italy,\\
INFN, Sezione di Lecce, Via Arnesano, I-73100 Lecce, Italy\\
$^*$E-mail: michele.maiorano@le.infn.com\\
}

\begin{abstract}
Even though Pulsar Timing Arrays already have the potential to detect the gravitational wave background by finding a quadrupole correlation in the timing residuals, this goal has not yet been achieved. Motivated by some theoretical arguments, we analyzed some advantages of including the millisecond pulsars within globular clusters, especially those in their cores, in current and future Pulsar Timing Array projects for detecting the gravitational waves emitted by an ensemble of supermassive black holes.
\end{abstract}


\bodymatter

\section{Introduction}
\label{sec:int}
Within the context of the Albert Einstein theory of General Relativity\cite{einstein1916}, gravitational waves (GWs) are space perturbations propagating at the speed of light in the vacuum. Due to the smallness of their amplitude, which in most cases is less than $10^{-16}$\footnote{This is a dimensionless quantity describing how a spatial dimension is stretched or squeezed by the passage of a GW.}, their detection is a challenging problem and, in fact, it took a hundred years since they were predicted for the first time. Nowadays, ground-based gravitational interferometers, such as those of the LIGO (Laser Interferometer Gravitational-Wave Observatory), KAGRA (Kamioka Gravitational Wave Detector) and VIRGO collaboration, are currently the only instruments capable of efficiently detecting GWs. Their sensitivity allows to detect high-frequency (i.e. in the range $10\text{--} 10^3$ Hz) GWs emitted by compact object systems, such as black hole binaries\cite{abbott2016}, neutron star binaries\cite{abbott2017} and neutron star-black hole binaries\cite{abbott2021}, in the final phase of the merger. The detection of low-frequency (i.e. in the frequency range $\simeq 10^{-5}\text{--} 1$ Hz) GWs is more challenging because it requires space-based gravitational interferometers such as LISA (Laser Interferometer Space Antenna), which is planned to be launched for $2034$ and consists of a constellation of three satellites, separated by about $1.5\times 10^6$ km, in a triangular configuration, orbiting with Earth around the Sun\cite{lisa2017,lisaweb}. However, ultra-low frequency (i.e. in the frequency range $\simeq 10^{-10}\text{--} 10^{-6}$ Hz) GWs, that may be generated by sources relevant for astrophysics and cosmology, such as supermassive black hole binaries (SMBHBs)\cite{rajagopal1995} and cosmic strings\cite{damour2001}, and form the gravitational wave background (GWB), are out of the range of any kind of gravitational interferometers and can only be detected by Pulsar Timing Arrays (PTAs)\footnote{For a quick graphic overview of the different GW sources and GW detectors, see Ref. \citenum{gwplotweb}.}.

PTAs constantly monitor the pulsed radio emission from the most stable isolated and binary millisecond pulsars across the sky to detect variation between the observed pulse time of arrival (ToA) and the one expected from the timing model. Such variation, referred to as timing residuals, may be GW-induced. Indeed, GWs can modify the distance between the Earth and the MSP, causing advances or delays in the pulse ToAs\cite{sazhin1978,detweiler1979}. As GWs, many other effects may induce timing residuals, so they must be distinguished from GWs. At present, the only way of doing this is by verifying that the cross-correlations between the timing residuals of each MSP pair follow the Hellings and Downs function, which is due to the quadrupole nature of GWs\cite{hellings1983}.

The main PTA collaborations are the European Pulsar Timing Array (EPTA), the Indian Pulsar Timing Array (InPTA), the North American Nanohertz Observatory for Gravitational Waves (NANOGrav), and the Parkes Pulsar Timing Array (PPTA), and all of them are part of the International Pulsar Timing Array (IPTA). These PTAs have timed about forty MSPs for more than ten years with an accuracy that should be sufficient, in principle, to detect gravitational waves. Some clues of the presence of a common red process compatible with a GWB was found, but there is still not a strong enough evidence for or against it\cite{arzoumanian2020,blasi2021,goncharov2021}. The reason behind that is yet unclear. Most likely, this could be due to the small number of MSPs available for PTAs, so continuing to add suitable MSPs should allow claiming detection, or to an insufficient observation time spawn, so it is essential to continue collecting data for many more years. Another option is that the standard GW search technique needs to be complemented by independent methods.

\section{Back to the Theory}\label{sec:bacthe}
The GW-induced timing residuals are described, in a compact form, by\footnote{For a complete theoretical description, see Ref. \citenum{maggiore2008b}.}
\begin{equation}\label{eq:timres}
r(t,\Omega^i)=\operatorname{Re}\left\lbrace\frac{h}{j\omega}\sum_{A=+,\times}\frac{1}{2}\frac{p^ip^je_{ij}^A}{(1+\Omega_ip^i)}\left[e^{j(\omega \tilde{t}-\alpha^A)}-e^{j(\omega \tilde{t} -\omega \tau(1-\Omega_i p^i)-\alpha^A)}\right]_0^t\right\rbrace
\end{equation}
where $h$, $\omega$, $A$, $e_{ij}^A$, and $\alpha^A$ indicate the GW amplitude, angular frequency, polarization state, polarization tensor, and the initial phase, respectively, $\tilde{t}$ is a mute variable, $\tau$ is the time distance between the Solar System Barycenter (SSB), $\Omega_i$ and $p^i$ indicate the direction versors of the GW source and the MSP, respectively. The two terms in the square brackets are known as Earth\footnote{The name ‘‘Earth'' is a legacy of early work on PTAs\cite{lommen2015}. In fact, since the Earth is not an inertial reference frame, the observer is assumed in SSB, whose position is known with great precision.} term and Pulsar term, and describe the metric in the proximity of the two objects\footnote{In this paper geometrical units c=G=1 have been adopted.}. Eq. \eqref{eq:timres} implies that, for an arbitrary MSP set, the Earth term is the same, while the Pulsar term changes. Since the latter term gives rise to an uncorrelated contribution in timing residuals, it is neglected in the standard GWB search\footnote{However, it is important to remark that the pulsar term may play an important role in the detection of GW emitted by single sources.}. This assumption is not entirely true if the MSP set is confined in a very small region, as in the case of globular clusters (GCs). Therefore, the GW-induced timing residuals of MSPs within GCs, especially those in their cores, should be strongly correlated, so considering these MSPs may offer a new instrument for GW detection.

\section{Simulated Environment}\label{sec:simsce}
The main target of the GW search with PTA is the GWBs produced by a hypothetical SMBHB population. Such a detection would give crucial information on the history of the Universe and, since that has not yet been achieved, it is worth considering improving PTAs by including MSPs within GCs.

\begin{figure}[H]
\centering
\includegraphics[scale=0.55]{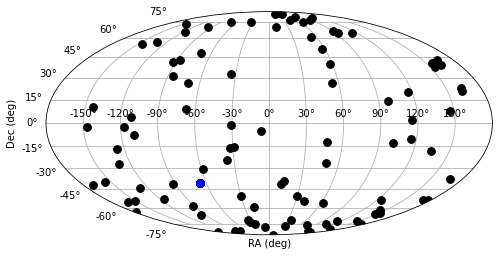}\hfill
\caption{Simulation of the SMBHB population. The position of $100$ simulated SMBHBs, randomly scattered all over the sky is shown. The black dots indicate the SMBHB coordinates (RA and Dec), while the blue dot indicates the simulated GC coordinates.}
\label{fig:skymap}
\end{figure}

\begin{figure}[H]
\centering
\subfloat[SMBHB chirp mass distribution.\label{pan:a}]{\includegraphics[scale=0.46]{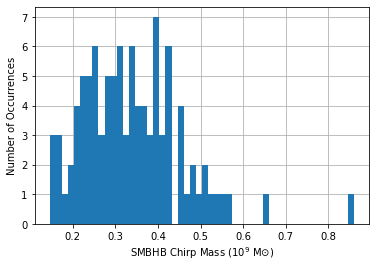}}
\subfloat[SMBHB distance distribution.\label{pan:b}]{\includegraphics[scale=0.46]{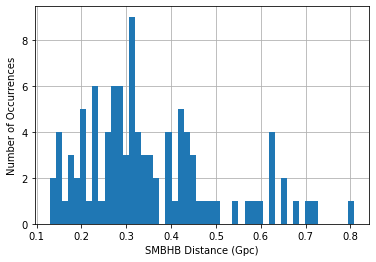}}\hfill

\subfloat[GW frequency distribution.\label{pan:c}]{\includegraphics[scale=0.46]{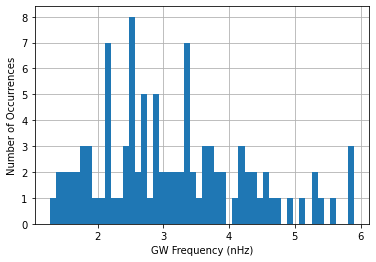}}
\subfloat[GW strain distribution.\label{pan:d}]{\includegraphics[scale=0.46]{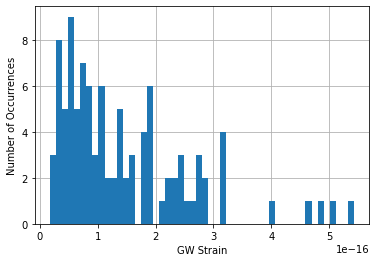}}\hfill
\caption{SMBHB parameter simulation. Panel \protect\subref{pan:a} shows the SMBHB chirp mass log-normal distribution, expressed in units of $10^9\,M_{\odot}$. Panel \protect\subref{pan:b} shows the SMBHB distance log-normal distribution, expressed in Gpc. Panel \protect\subref{pan:c} shows the GW frequency log-normal distribution, expressed in nHZ. Panel \protect\subref{pan:d} shows the GW strain distribution, expressed in units of $10^{-16}$. On the vertical axis of each panel is plotted the number of occurrences.}
\label{fig:smbsim}
\end{figure}

\begin{figure}[H]
\centering
\includegraphics[scale=0.70]{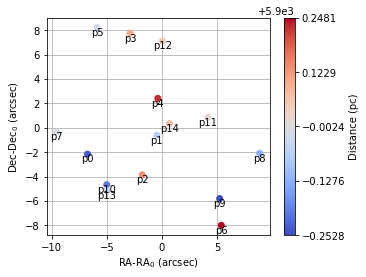}\hfill
\caption{Simulation of the MSP set. The position of $15$ simulated GC core MSPs, randomly scattered within a GC core, whose distance and angular radius are $5.9$ kpc and $0.16$ arcmin, respectively, is shown. The dots indicate the MSP coordinates (RA and Dec), with respect to the GC center (RA$_0$ and Dec$_0$), expressed in arcsec, while the dot colors indicate, as described by the color bar, the MSP distance with respect to the GC core center, expressed in pc (the MSP distance with respect to the SSB can be obtained by summing the values expressed by the labels of the color bar with the quantity on its top).}
\label{fig:glomsp}
\end{figure}

\begin{figure}[H]
\centering
\subfloat[Simulated $30$ years GW-induced timing residuals. \label{pan:30}]{\includegraphics[scale=0.55]{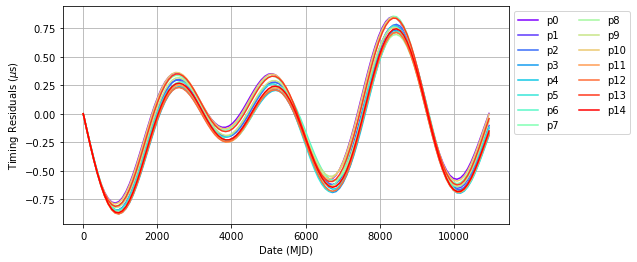}}

\subfloat[Simulated $10$ years GW-induced timing residuals. \label{pan:10}]{\includegraphics[scale=0.55]{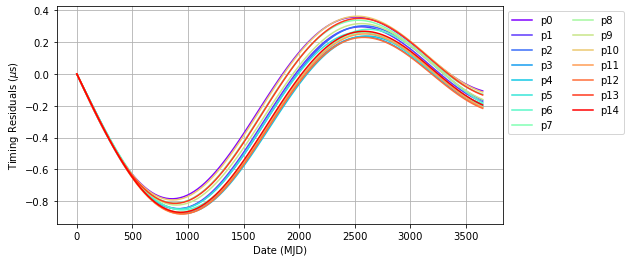}}
\caption{GW-induced timing residuals. Panel \protect\subref{pan:30} shows the simulated GW-induced timing residuals corresponding to each GC core MSP due to the overall contribution of the GW emission from each SMBHB, plotted over a time interval of $30$ years. Panel \protect\subref{pan:10} shows the GW-induced timing residuals, plotted over a time interval of only $10$ years. In both panels, the colored curves indicate the simulated GW-induced timing residuals, and have a different color for each GC core MSP, as described by the legend. On the horizontal axis of both panels the date measured from the arbitrary beginning of the simulated ToA observation, expressed in MJD, is plotted. On the vertical axis of both panels the timing residuals, expressed in $\mu$s, are plotted.}
\label{fig:timres}
\end{figure}

We simulated the GW-induced timing residuals on GC core MSPs due to an SMBHB population. We arbitrarily assumed that the SMBHB chirp mass, distance, and GW emission frequency follow a log-normal distribution\footnote{The properties of the SMBHB population are still a matter of debate. However, this paper aims to explain how PTAs can be improved by including MSPs within GCs, so we took the results in Ref. \citenum{tucci2017,celoria2018,sanchis2021} just as a guide for building SMBHB distributions compatible with the GW-induced timing residuals observable by PTAs.} and are in the ranges $10^8\text{--} 10^9$ M$_\odot$, $10^8\text{--} 10^9$ pc, and $10^{-9}\text{--} 10^{-8}$ Hz, respectively, and that the SMBHB right ascension and declination follow a uniform distribution (see Fig. \ref{fig:skymap}).
The SMBHB strain is, in good approximation, given by:
\begin{equation}\label{eq:smbstr}
h=\sqrt{\frac{32}{5}}\frac{\mathcal{M}^{5/3}}{D}\left(\frac{2\pi f}{1+z}\right)^{2/3}
\end{equation}
where $\mathcal{M}$, $D$, $z$, and $f$ indicate the SMBHB chirp mass, distance, cosmological redshift, and GW emission frequency, respectively. Eq. \eqref{eq:smbstr} allows to determine the GW strain distribution for our SMBHB population (see Fig. \ref{fig:smbsim}).
We also assumed that the GC core MSP distance\footnote{In this paper, MSP distance refers to the distance with respect to the diametrical plane passing through the GC center and orthogonal to the line-of-sight.}, right ascension, and declination with respect to the GC center follow a uniform distribution\footnote{In accordance with what is described by the GC core mass-density distribution\cite{binney1987}.}. To get a realistic result, we took the GC core distance, angular radius, and MSP number of the Terzan 5 GC since it is one of the most populated by MSPs\cite{cadelano2018,atnfweb,freireweb} (see Fig.\ref{fig:glomsp}).
We finally calculated the simulated GW-induced timing residuals corresponding to each GC core MSP due to the overall contribution of the GW emission from each SMBHB (see Fig. \ref{fig:timres}).

\section{Results}\label{sec:resdis}
The simulation results (see Fig. \ref{fig:timres}) confirm what has been deduced in Sec. \ref{sec:bacthe}. In the PTA frequency band, the simulated GW-induced timing residual corresponding to each GC core MSP due to the overall contribution of the GW emission from each SMBHB appear to be strongly correlated. This property can be very useful in the GW search. Observing such correlation for the GC core MSPs would lead to a strong evidence for the GWB, especially if supported by stronger hints of a common red process compatible with a GWB. GC core MSPs can also be helpful for determining the quadrupolar nature of the cross-correlations between the timing residuals of each MSP pair. Indeed, the Hellings and Downs function has a global maximum when the angular distance between an MSP pair is null. Of course, it is unlikely to have an aligned MSP pair by chance, but this frequently happens in GCs due to physical reasons. Therefore, GCs offer the unique possibility of obtaining cross-correlation data for low angular separation MSPs, which can be very important because they might challenge or confirm the prediction of the Hellings \& Downs function. 

\section{Conclusion}\label{sec:con}
The proximity between the MSPs lying in a given GC, especially those in the GC core, can be exploited to get unique information about GWs and, for this reason, it is worth considering improving PTAs by including them.

It is important to note that, in the presented analysis, it has been considered an ideal situation to highlight some of the advantages of including MSPs within GCs in PTAs. On the other hand, this class of MSPs is currently not timed with a precision adequate to detect GW-induced timing residual\cite{damico2009,freire2017}.

Nevertheless, we expect that to change soon, thanks to new detectors, such as the Five hundred meter Aperture Spherical Telescope (FAST) \cite{nan2011}, the MeerKAT radio telescope \cite{bailes2020}, and the Square Kilometre Array (SKA) \cite{weltman2020}. SKA, in particular, will play a crucial role in ultra-low frequency GW detection. With its unprecedented sensitivity, it will be possible to time significantly better the GC MSPs, opening the possibility of adopting the strategy proposed in this paper, which is a summary of Ref. \citenum{maiorano2021}.

\section*{Acknowledgments}\label{sec:ack}
We warmly acknowledge Andrea Possenti, of the Istituto Nazionale di Astrofisica (INAF), for many useful discussions. We also acknowledge the support by the Theoretical Astroparticle Physics (TAsP) and Euclid projects of the Istituto Nazionale di Fisica Nucleare (INFN).

\bibliographystyle{ws-procs961x669}
\bibliography{ws-pro-sample.bib}

\end{document}